\documentclass[review]{elsarticle}
\usepackage{epstopdf}
\usepackage{lineno,hyperref}

\journal{Journal of \LaTeX\ Templates}
\usepackage{graphicx}
\usepackage{amssymb,bm,mathrsfs,bbm,amscd}

\begin{document}

\begin{frontmatter}

\title{A study of aging effects of barrel Time-Of-Flight system in the BESIII experiment\tnoteref{mytitlenote}}
\tnotetext[mytitlenote]{Supported in part by National Natural Science Foundation of China (U1232201, 11575225, 11605220, 11675184, 11735014), National Key Basic Research Program of China (2015CB856700), the CAS center for Excellence in Particle Physics.}

\address[mymainaddress]{Institute of High Energy Physics, Chinese Academy of Sciences, Beijing 100049, China}
\address[mysecondaryaddress]{University of Chinese Academy of Sciences, Beijing 100049, China}
\address[mytertiaryaddress]{State Key Laboratory of Particle Detection and Electronics, Beijing 100049, China}
\cortext[mycorrespondingauthor]{Corresponding author}

\author[mymainaddress,mysecondaryaddress]{Huan-Huan Liu\corref{mycorrespondingauthor}}
\ead{hhliu@ihep.ac.cn}
\author[mymainaddress,mysecondaryaddress]{Sheng-Sen Sun\corref{mycorrespondingauthor}}
\ead{sunss@ihep.ac.cn}
\author[mymainaddress,mysecondaryaddress]{Shuang-Shi Fang}
\author[mymainaddress,mytertiaryaddress]{Zhi Wu}
\author[mymainaddress,mytertiaryaddress]{Hong-Liang Dai}
\author[mymainaddress,mysecondaryaddress,mytertiaryaddress]{Yue-Kun Heng}
\author[mymainaddress,mysecondaryaddress]{Ming Zhou}
\author[mymainaddress]{Zi-Yan Deng}
\author[mymainaddress]{Huai-Min Liu}

\begin{abstract}
The Time-Of-Flight system consisting of plastic scintillation counters plays an important role for particle identification in the BESIII experiment at the BEPCII double ring $e^+e^-$ collider. Degradation of the detection efficiency of the barrel TOF system has been observed since the start of physical data taking and this effect has triggered intensive and systematic studies about aging effects of the detector. The aging rates of the attenuation lengths and relative gains are obtained based on the data acquired in past several years. This study is essential for ensuring an extended operation of the barrel TOF system in optimal conditions.
\end{abstract}


\end{frontmatter}


\section{Introduction}

The Beijing Spectrometer (BESIII) detector~\citep{bes3} was designed to study physics in the $\tau$-charm energy region at the high luminosity Beijing Electron Positron Collider (BEPCII)~\citep{bepc2design,bepc}.
The Time-Of-Flight (TOF) system based on plastic scintillation counters plays a key role for particle identification in the BESIII experiment, especially for $K/\pi$ separation; it can also provide fast trigger signals.
The TOF system consists of a double layer barrel and two single layer end caps~\citep{bes3,tof1}.
In the global BESIII coordinate system, the beam direction is chosen as the $\hat{z}$ axis and the polar angle coverage of the barrel TOF is $|\cos\theta|<0.83$, while that of the end cap is $0.85<|\cos\theta|<0.95$.
The end cap Time-Of-Flight detector was upgraded with multi-gap resistive plate chamber (MRPC) technology in summer of 2015~\citep{mrpc1,mrpc2,mrpc3,mrpc4}.
In this article, the investigation of the aging effects on the performance of the scintillation counters of the barrel TOF system of BESIII detector is described.

\begin{figure}[!htbp]
\begin{center}
\includegraphics[width=6cm]{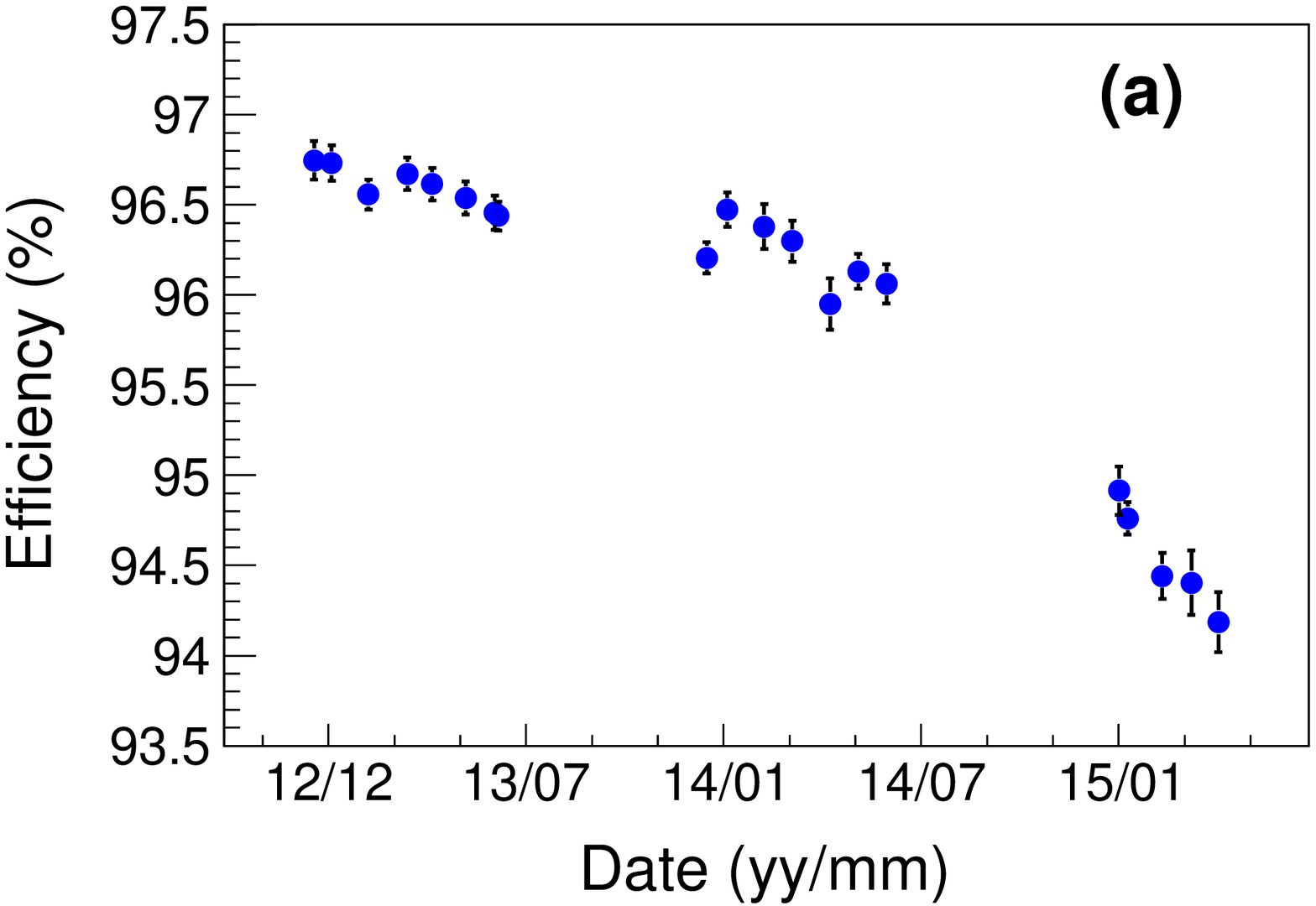}
\includegraphics[width=6cm]{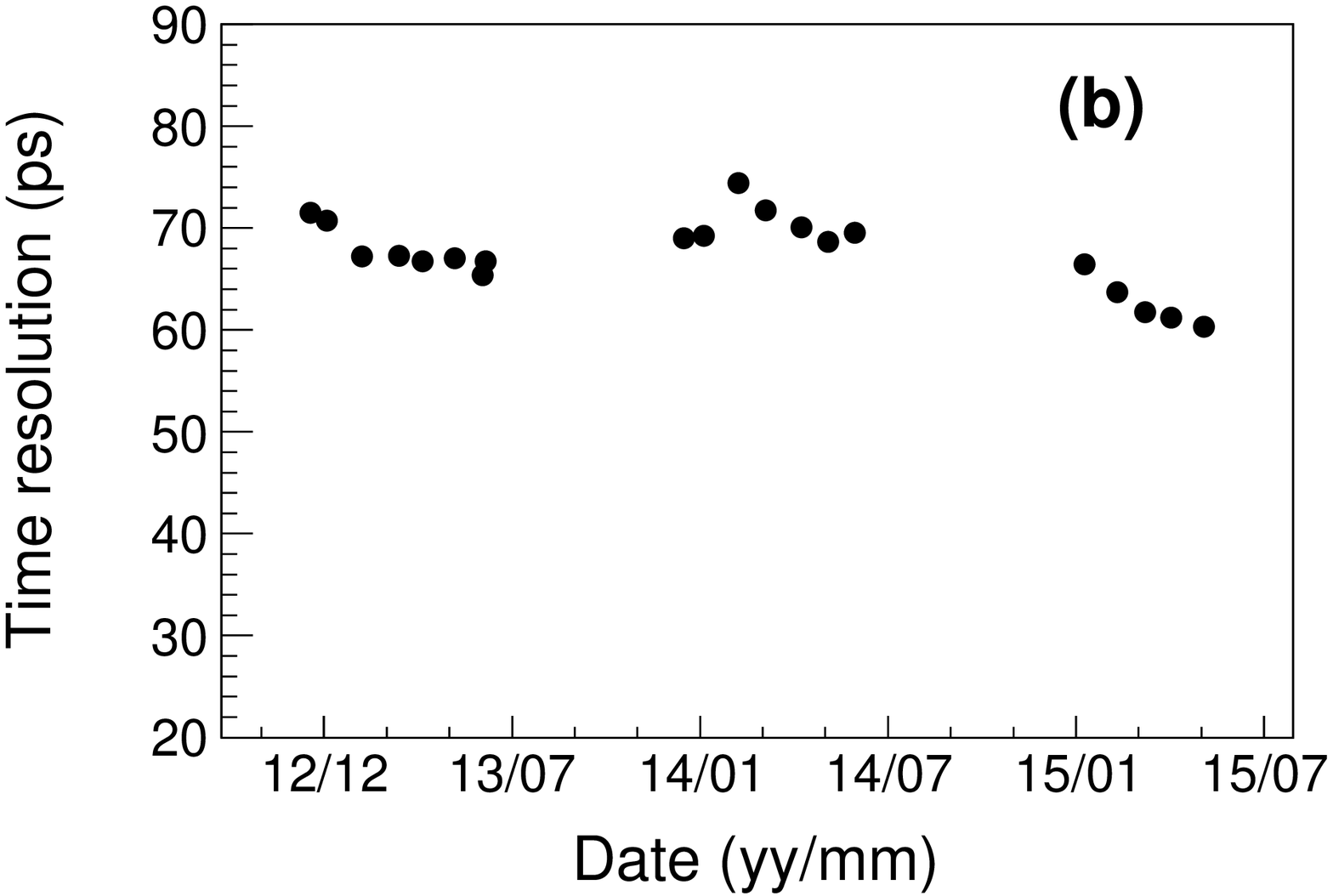}
\caption{(a) Time dependence of detection efficiency for a typical scintillation counter of the barrel TOF system; (b) Time dependence of overall time resolution of the barrel TOF system of BESIII. }
\label{eff}
\end{center}
\end{figure}

The physics data taking of BESIII detector was began in 2009.
The detection efficiency of barrel TOF system is found to be decreased with time since then, and no obvious deterioration of overall time resolution, not ``intrinsic'' time resolution, is observed.
The time dependences of detection efficiency of electrons or positrons in Bhabha events for a typical scintillation counter and overall time resolution of the barrel TOF system are shown in Fig.~\ref{eff} (a) and (b) respectively.
The detection efficiency is defined as $\varepsilon={n_{\Delta{t}}}/{n_{track}}$, $n_{track}$ is the number of the main drift chamber(MDC) tracks extrapolated to barrel TOF, and $n_{\Delta{t}}$ is the number of MDC tracks with their time differences between measured and expected times less than a specific value, such as 1 ns, in order to suppress the background.
The reduction of the number of output signals from the high threshold discriminators caused by the aging effect of scintillation counters and photomultiplier tubes predominately contributes to the performance degradation of the TOF detector.
The high voltages (HVs) of the photomultiplier tubes (PMTs) of the barrel TOF system were increased twice in 2012 and 2016, the reduced output signals were amplified and the efficiency-drops were recovered.
There are several factors contribute to the overall time resolution of the barrel TOF system at BESIII, including the ``intrinsic'' time resolution, uncertainties of time and length of beam bunch, uncertainty of hit position and the expected time, time resolution of electronics and uncertainty caused by the time walk~\citep{bes3}.
Since the contribution of each term was different depends on the performance of the detector and accelerator, we could not arbitrarily draw a conclusion that the ``intrinsic'' time resolution did not degrade.
The two times increasements of HVs enhanced the quantum efficiencies of the PMTs, and were helpful for the stability of the number of photoelectrons and the ``intrinsic'' time resolution.

Time dependent degradation of detector performance is a phenomenon which has been observed for a long time.
Serval investigations of aging effects of the scintillation counters and PMTs were applied by several particle physics experiments, such as the attenuation length aging effect of Belle TOF system~\citep{belle-tof1, belle-tof2} and the aging of light yield in the CDF II scintillation counters~\citep{cdf-ly1, cdf-ly2}.
With a view to the importance of particle identification and its maintenance over the whole lifetime of the experiment, an intensive and systematic study of aging effects onto the TOF system is essential for making long-term forecasts.

\section{The barrel Time-Of-Flight system}

The barrel TOF system has two layers of staggered scintillation counters mounted on the outer surface of the MDC.
Each layer has 88 plastic scintillation counters~\citep{tof2,tof3,tof4} read out by fine-mesh photomultiplier tubes.
The barrel counters are Bicron 408 scintillator with a trapezoidal cross-section, and the length and thickness are 2300 mm and 50 mm respectively.
Two PMTs (Hamamatsu R5924) are attached to the two ends of a counter and coupled by a 1 mm thick silicone pad (BC\-634A).
The length of the PMT is 50 mm, and the total length of PMT assembly is 103 mm, including the base and preamplifier.
The TOF readout system~\citep{tof5} consists of preamplifiers mounted in the PMT bases, cables, signal time and amplitude measurement circuits and a laser calibration system.
A dual threshold discriminator scheme, designed to reduce noise while maintaining low time walk, is used for time digitization.
The outputs from the low threshold discriminators of the two ends are used to start the precision time-to-digital conversion (TDC) process and are put into coincidence with the high threshold discriminator outputs in order to reduce background rates.
The signal amplitude measurement circuit is based on the charge-to-time conversion (QTC) principle.
A more complete description of the Time-Of-Flight system employed at BESIII, its design and arrangement can be found in the articles~\citep{bes3,tof1}.
Since TOF usually operates in a radiation environment, the radiation damage, including light transmission, light output, excitation and emission spectra, of Bicron 408 and other two types of plastic scintillators had been experimentally studied~\citep{dose} in the research and design stage of TOF system at BESIII.
The performance of fine-mesh Hamamatsu R5924 PMT had also been investigated in a strong magnetic field up to 1T~\citep{PMT}, and an exponential dependence of the relative gain on the working voltage was observed.

\section{Attenuation length measurements}

The plastic scintillators exploit the ionization produced by incident charged particles to generate optical photons.
Organic scintillators, including plastic scintillators, do not respond linearly to the ionization density.
A semi-empirical model by Birks describes the light-output degradation at high ionization density~\citep{birks}:
\begin{eqnarray}
\label{birks}
\frac{d\mathscr{L}}{dx}=\mathscr{L}_{0}\frac{dE/dx}{1+{k}_{B}dE/dx},
\end{eqnarray}
where $\mathscr{L}$ is the luminescence, $\mathscr{L}_0$ is the luminescence at low specific ionization density, $dE/dx$ is the ionisation energy loss in $MeV/(g/cm^2)$ and ${k}_{B}$ is Birks' constant, which is characteristic of the scintillation material.
The produced photons are then propagated by total internal reflection, refraction and absorption in the scintillator until they arrive at the PMTs located at the ends of the scintillator and produce a signal pulse.
This light attenuation is described as an exponential function with an attenuation length $\lambda$ which is defined as the length reducing the light signal by a factor of $e$, and attenuation length is crucial to achieve a high light collection efficiency for scintillation counters.
For the barrel TOF, the measured signal pulse height(charge) QTCs at the two ends of a scintillator, $q_1$ and $q_2$ can be expressed as:
\begin{eqnarray}
\label{charge}
&&{q_1}={A_1}{\cdot}\frac{q_0}{\sin\theta}{\cdot}\exp{\left(-\frac{l/2-z}{\lambda}\right)},\nonumber\\[2mm]
&&{q_2}={A_2}{\cdot}\frac{q_0}{\sin\theta}{\cdot}\exp{\left(-\frac{l/2+z}{\lambda}\right)},
\end{eqnarray}
where the subscripts 1 and 2 represent the readout channels in the directions along the positron beam and the opposite, respectively;
$A_1$ and $A_2$ are the relative gains, including the contributions from the light yields of the scintillation bars, the quantum efficiencies (QEs), photon-electron collection efficiencies (CEs), multiplication gains of PMTs and preamplifier gains for the two readout channels respectively;
$\theta$ is the polar angle of the incident particle with respect to the direction of the positron beam;
$l$ is the total length of the scintillator bar;
$z$ is the extrapolated hit position on the internal surface of the scintillator bar along the beam direction based on the particle trajectory~\citep{extrapolation};
$q_0$ is the normalized pulse height at $z=0$ and ${\theta}={90}^{\circ}$, which is proportional to the luminescence of a vertical incident particle.

\begin{figure}[!htbp]
\begin{center}
\includegraphics[width=6cm]{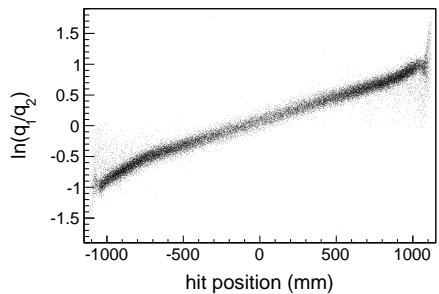}
\caption{\label{qvsz} Scattering plot of $\ln(q_1/q_2)$ versus hit position $z$. }
\end{center}
\end{figure}

From Eq.~\ref{charge}, it is easily to deduce the following equation
\begin{eqnarray}
\label{latten_e}
\ln\left(\frac{q_1}{q_2}\right)=\ln{\left(\frac{A_1}{A_2}\right)}+\frac{2{\cdot}z}{\lambda}.
\end{eqnarray}
The scattering plot of $\ln(q_1/q_2)$ versus hit position $z$ of incident charged particles is shown in Fig.~\ref{qvsz}.
The attenuation length $\lambda$ and relative gain ratio $A_1/A_2$ can be extracted by means of a linear fitting to $\ln(q_1/q_2)$ versus hit position $z$ using real data.
The nonlinearity behavior close to two ends of the scintillator bar is caused by the saturation of QTC measurements and edge effects.

The attenuation length dependence as a function of time is expressed as:
\begin{eqnarray}
\label{life}
{\lambda}(t)={\lambda}_0e^{-{{\alpha}_{\lambda}}{\cdot}t},
\end{eqnarray}
where ${\lambda}_0$ is the attenuation length at the beginning of measurements; ${\lambda}(t)$ is the attenuation length at time $t$; $t$ is the time; ${\alpha}_{\lambda}$ is the attenuation length aging constant for the counter.

\begin{figure}[!htbp]
\begin{center}
\includegraphics[width=6cm]{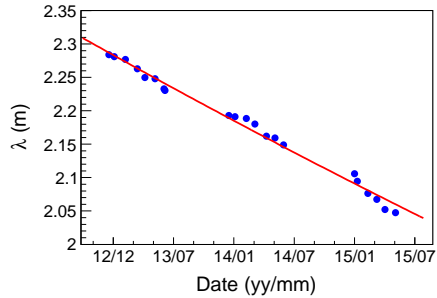}
\caption{\label{latten_f} Time dependence of attenuation length for a typical scintillation counter of the barrel TOF system. }
\end{center}
\end{figure}

\begin{figure}[!htbp]
\begin{center}
\includegraphics[width=6cm]{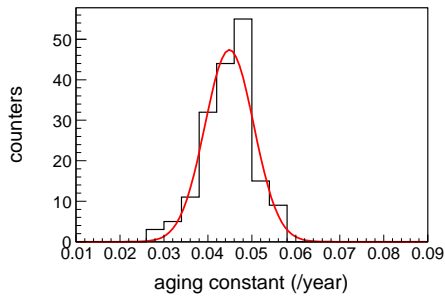}
\caption{\label{suml} Distribution of the individually calculated attenuation length aging constant for each scintillation counter together with a Gaussian distribution fitted to the data. }
\end{center}
\end{figure}

Data samples of electrons and positrons from Bhabha events which are selected using online event filtering algorithm~\citep{evtfilter} are employed for the measurements of attenuation lengths.
In the course of several years, measurements of attenuation length of each scintillator counter have been performed periodically using the method summarized in Eq.~\ref{latten_e}.
The plot of time dependence of the attenuation length for a typical scintillation counter of the barrel TOF system, together with an exponential function from Eq.~\ref{life}, is shown in Fig.~\ref{latten_f}.
The attenuation length aging constant is measured individually for each scintillation counter.
The resulting distribution of attenuation length aging constants for 176 scintillation counters of barrel TOF system is shown in Fig.~\ref{suml}, which is approximately described by a Gaussian function.
The average experimental value of attenuation length aging constants ${\alpha}^{ave}_{\lambda}=0.045/$year, corresponds to a degradation annual rate of about $4.4\%$.

\section{Relative gains}

From 2012 to 2015, the center-of-mass energies of physics data taking were 2-4.6 GeV, and correspondingly the ${\beta}{\gamma}$s of electrons and positrons in Bhabha events were between 2000 and 4600.
Since the relativistic rise of the energy loss saturates in this ${\beta}{\gamma}$ range as described by the ``Bethe equations''~\citep{betheformula}, the differences of mean energy loss per unit length for Bhabha events taken at different energies is negligible.
That means that the luminescence intensity or normalized pulse height $q_0$ for Bhabha events at different energies would remain constant.
In Eq.~\ref{charge}, the relative gains $A_1$ and $A_2$ have contributions from the light yield of the scintillator bars, efficiencies and multiplication gains of PMTs and preamplifier gains of the two readout channels.
Presently, it is extremely difficult to access the scintillation counters in their positions inside the BESIII detector.
Therefore it is impossible to extend the investigation of the aging effects to a large number of scintillator bars, PMTs and electronics preamplifiers.
However, since the degradation of the measured signal pulse height is expected to be accompanied by the aging effect from the relative gains of each part of this sequential chain, the investigation of the overall aging effect is performed instead of separate investigations of aging effects of scintillator bars, PMTs and electronics channels.

The products between the relative gains at the two ends of one scintillator bar and the normalized pulse height can be derived from Eq.~\ref{charge},
\begin{eqnarray}
\label{normalq}
&&{A_1}{\cdot}{q_0}={q_1}{\cdot}{\sin\theta}{\cdot}{\exp{\left(\frac{l/2-z}{\lambda}\right)}},\nonumber\\[2mm]
&&{A_2}{\cdot}{q_0}={q_2}{\cdot}{\sin\theta}{\cdot}{\exp{\left(\frac{l/2+z}{\lambda}\right)}}.
\end{eqnarray}

\begin{figure}[!htbp]
\begin{center}
\includegraphics[width=6cm]{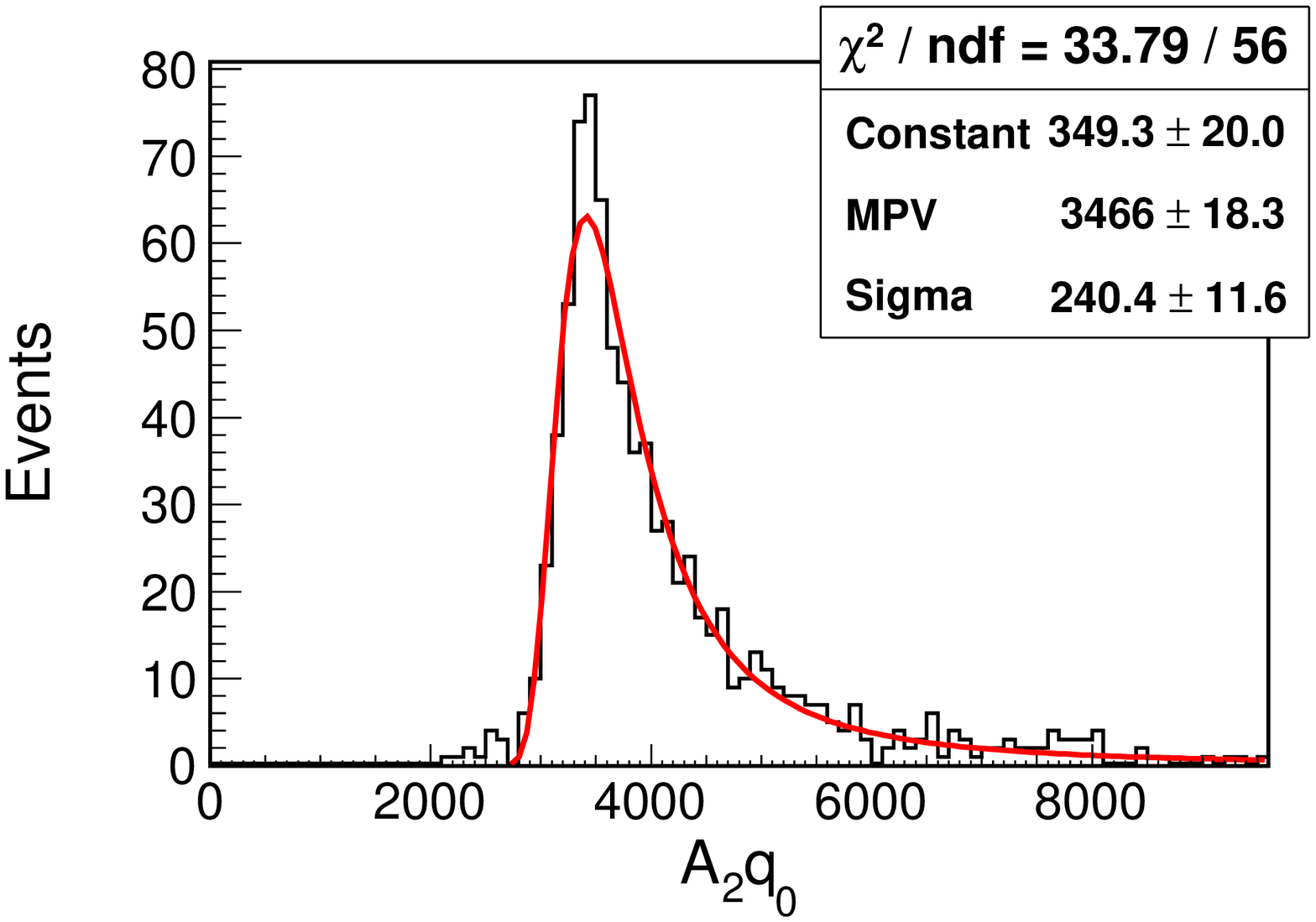}
\caption{\label{aq0} The distribution between the product of relative gain and normalized pulse height of a typical electronics channel. }
\end{center}
\begin{center}
\includegraphics[width=6cm]{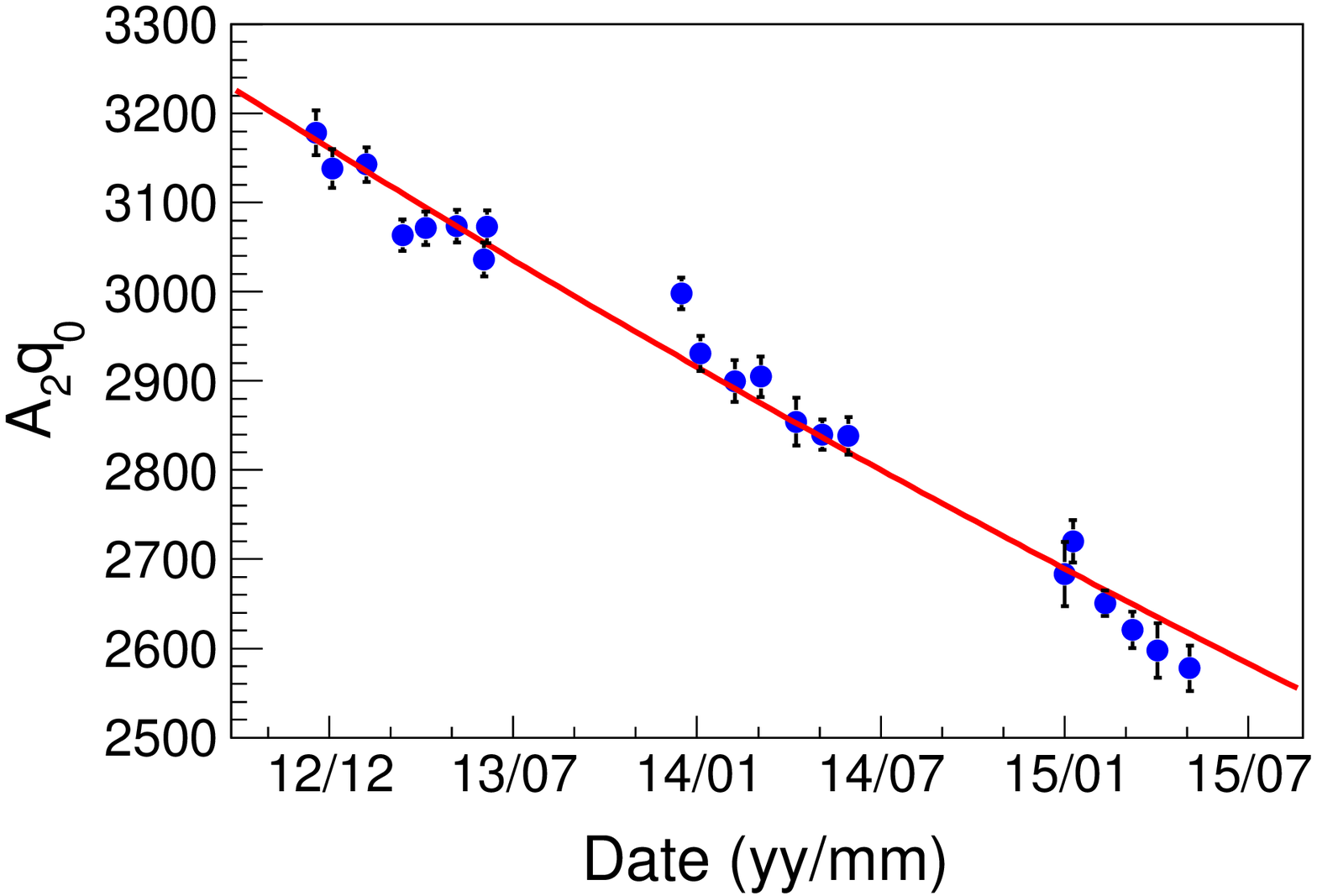}
\caption{\label{aq0time} The distribution of time dependence of the most probable value of the product between the relative gain and normalized pulse height for a typical electronics readout channel. }
\end{center}
\begin{center}
\includegraphics[width=6cm]{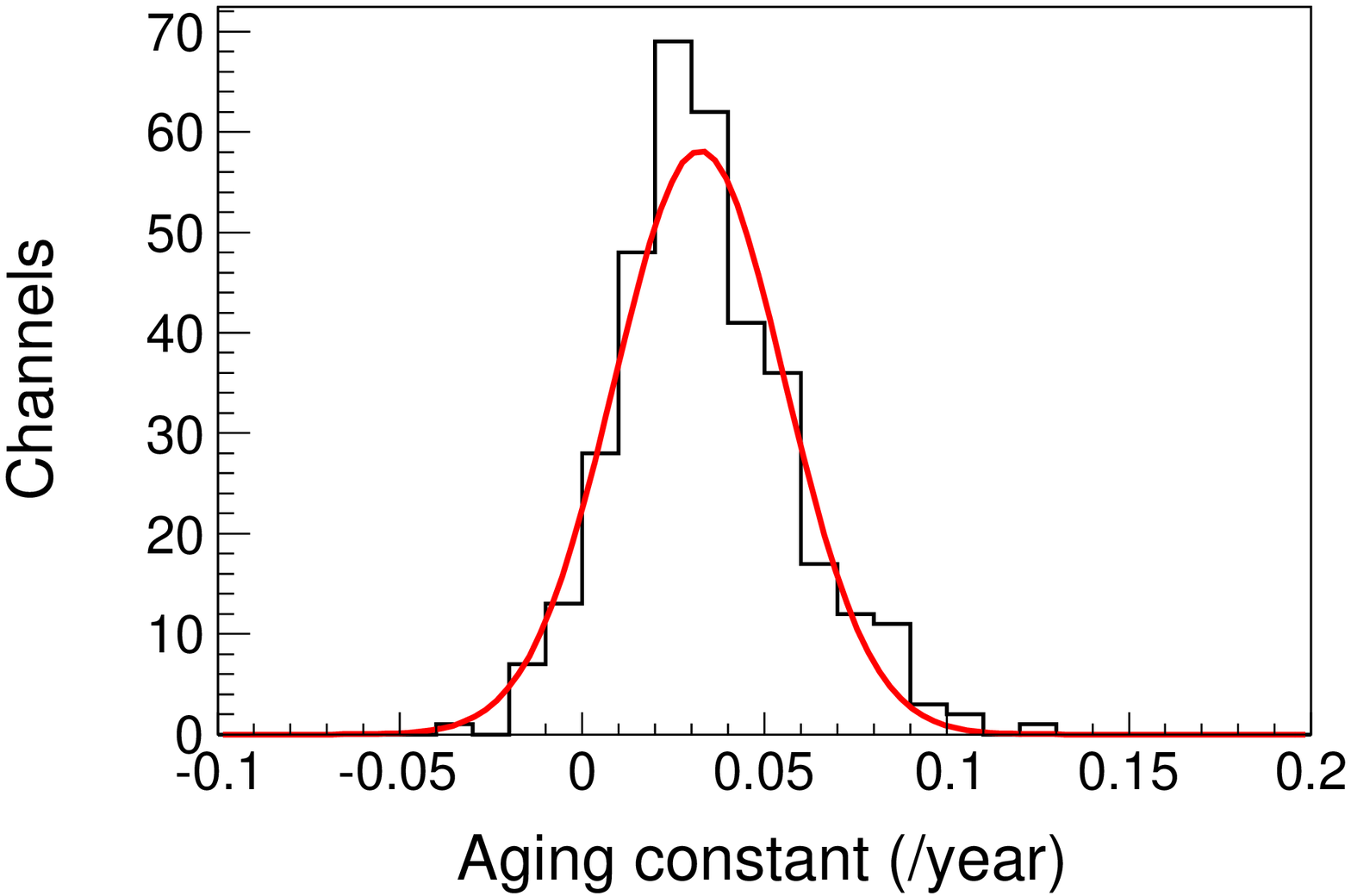}
\caption{\label{sumgain} A Gaussian fit is superimposed to the distribution of the individually calculated aging constants for each readout channel. }
\end{center}
\end{figure}

Using the same data sample of Bhabha events already used for the study of the attenuation length, the distribution between the products of the relative gain and the normalized pulse height of a typical electronics channel is shown in Fig.~\ref{aq0}, together with the best approximation using the Landau function~\citep{landau}.

The plot of the time dependence of the most probable value (MPV) of the product between the relative gain and the normalized pulse height of a typical channel is shown in Fig.~\ref{aq0time}.
Since the ideal normalized pulse height $q_0$ approximates to a constant without time dependence, an exponetial behaviour very similar to the one already used in Eq.~\ref{life} is expected and used to fit the distribution:
\begin{eqnarray}
\label{gainlife}
{A}(t)={A_0}e^{-{{\alpha}_A}{\cdot}t}
\end{eqnarray}
where $A_0$ is the relative gain at the beginning of measurements; $A(t)$ is the relative gain at time $t$, $t$ and ${\alpha}_A$ are the time and the relative gain aging constant for the electronics readout channel respectively.

For each electronics readout channel, the relative gain aging constant is obtained by fitting the data with Eq.~\ref{gainlife}.
The distribution for the aging constants of the relative gains for 352 readout channels of barrel TOF system is shown in Fig.~\ref{sumgain}, and this distribution is fitted using a Gaussian function.
The average experimental value for the aging constant of the relative gains ${\alpha}^{ave}_A=0.032 /$year, corresponds to degradation annual rate of about $3.2\%$.

\section{Estimation of lifetime of photomultiplier tubes}

The BESIII detector is configured around a 1T superconducting solenoid (SSM), considered to be the optimum for precise momentum measurements for charged tracks in the $\tau-$charm energy region.
Other two superconduction quadrupoles (SCQs) are inserted in the conical shaped MDC end caps as close as possible to the interaction point.
In this region less secondary electrons would hit the dynodes and as a consequence reduce the multiplication gain.
The fine-mesh PMTs R5924 by Hamamastu have a high resistivity against magnetic field because of their 19 stages of fine-mesh dynodes, which has many apertures to let electrons go through.
A fast preamplifier with a gain of 10 is installed in the PMT base to boost signals and extend the lifetime of the PMT.

Since the start of the engineering run of BEPCII in collision mode in 2008 to 2015, the total running time is about 15869 hours.
The average event rate is around 300 kHz with the events satisfying the TOF trigger condition, that is number of hits in the barrel TOF equal or larger than 1.
For each PMT, the average hit rate is about 0.04 per event and the most probable value of charge distribution is about 12 pC; these numbers are obtained using a random trigger data sample.
The integrated amount of output charge from individual PMT anode is 8.2 C in these 7 years.
This is a conservative estimation since small signals which could not be over threshold have no record in the raw data.
The magnitude of integrated charge corresponding to a 7 year time duration is far smaller than the limit of lifetime of PMTs, estimated to be around 360 C.

\section{Summary}

Investigation shows that the degradation of detection efficiency of barrel TOF system is related to aging effects in the detector, and that no significant deterioration in overall time resolution is observed since the start of physics data taking of BESIII.
Therefore we performed a study on the aging rates of the attenuation length and relative gain using Bhabha events taken in the period of 2012-2015.
An approximate estimation of the lifetime of the PMTs shows they should be safe under the present irradiation circumstance over the long term.
These numbers allow us to forecast the detection efficiency and make an appropriate plan to ensure the detector operation in optimal conditions.
The high voltages of the photomultiplier tubes of the barrel TOF system have been increased twice since 2009.
The relationship between the multiplication gain and PMT high voltage could be described using an exponential function and the gain will become saturated with enough high HV.
Except increasing the HV of PMTs, decreasing the thresholds of electronics is another choice for the recovery of the detection efficiency.
Some experiments about the tuning of the threshold are expected to provide a quantitative description of this behaviour.
The relative gain includes the contributions of light yield of the scintillator bar, efficiencies and gain of PMT and preamplifier gain of electronics channel.
In this study, the aging effect of the relative gain is considered as a cumulative result of all these contributions.
Individual investigations of aging properties of scintillation counters, PMTs and electronics would lead to an improved understanding of aging phenomena and enable to extend the operational period, therefore a further study is still necessary to obtain reliable long term forecasts, such as 8-10 years.
Thus a detailed proposal for studying the light yield of the scintillation counters, the quantum efficiency, photon-electron collection efficiency and multiplication gain of PMT and preamplifier gain for the readout channel is now in preparation.


\bibliography{mybibfile}

\begin{thebibliography}{10}
\expandafter\ifx\csname url\endcsname\relax
  \def\url#1{\texttt{#1}}\fi
\expandafter\ifx\csname urlprefix\endcsname\relax\def\urlprefix{URL }\fi
\expandafter\ifx\csname href\endcsname\relax
  \def\href#1#2{#2} \def\path#1{#1}\fi

\bibitem{bes3}
{BESIII Collaboration (M. Ablikim, et al.)}, {Design and Construction of the
  BESIII Detector}, Nucl. Instrum. Meth. {\bf{A 614}} (2010) 345--399.

\bibitem{bepc2design}
{BES Collaboration (J. Z. Bai, et al.)}, {The BES upgrade}, Nucl. Instrum.
  Meth. {\bf{A 458}} (2001) 627--637.

\bibitem{bepc}
{BESIII Collaboration (M. Ablikim, et al.)}, {Performance of the BEPC and
  progress of the BEPCII}, Proceedings of APAC (2004) p. 15--19.

\bibitem{tof1}
{Y. K. Heng, et al.}, {The Progress of TOF on BESIII}, IEEE Nucl. Sci. Symp.
  Conf. Rec. (2007) 53--57.

\bibitem{mrpc1}
{X. Z. Wang, et al.}, {The cosmic ray test of MRPCs for the BESIII ETOF
  upgrade}, Eur. Phys. J. {\bf{C 76}}~(4) (2016) 211.

\bibitem{mrpc2}
{X. Z. Wang, et al.}, {The upgrade system of BESIII ETOF with MRPC technology},
  JINST {\bf{11}}~(08) (2016) C08009.

\bibitem{mrpc3}
{Zhi Wu, et al. }, {First results of the new endcap TOF commissioning at
  BESIII}, JINST {\bf{11}}~(07) (2016) C07005.

\bibitem{mrpc4}
{X. Z. Wang, et al. }, {BESIII ETOF upgrade readout electronics commissioning},
  Chin. Phys. {\bf{C 41}}~(1) (2017) 016103.

\bibitem{belle-tof1}
{M. Jones, et al. }, {Calibration and performance of the Belle TOF system},
  Belle Note (2003) 596.

\bibitem{belle-tof2}
{M. Jones, et al. }, {Performance of the Belle TOF system 2000-2009}, Belle
  Note (2009) 1146.

\bibitem{cdf-ly1}
{A. Artikov, et al. }, {On the aging of the scintillation counters for RUN II
  Muon System at CDF}, Nucl. Instrum. Meth. {\bf{A 579}} (2007) 1122--1134.

\bibitem{cdf-ly2}
{A. Artikov, et al. }, {The loss of light yield with time in the CDF II
  scintillation counters}, Nucl. Instrum. Meth. {\bf{A 672}} (2012) 46--51.

\bibitem{tof2}
{Chong Wu, et al}, {The timing properties of a plastic time-of-flight
  scintillator from a beam test}, Nucl. Instrum. Meth. {\bf{A 555}} (2005)
  142--147.

\bibitem{tof3}
{Z. J. Sun, et al.}, {Beam Test for a 1:1 Module of Time of Flight Counter of
  BESIII}, High Energy Phys. Nucl. Phys. (in Chinese) {\bf{29}}~(10) (2005)
  933--937.

\bibitem{tof4}
{S. H. An, et al.}, {Testing the time resolution of the BESIII end-cap TOF
  detectors}, Measur. Sci. Tech. {\bf{17}} (2006) 2650--2654.

\bibitem{tof5}
{S. B. Liu, C. Q. Feng, Qi An, Y. K. Heng, S. S. Sun}, {BES III time-of-flight
  readout system }, IEEE Trans. Nucl. Sci. {\bf{57}} (2010) 419--427.

\bibitem{dose}
{Zhao Li, et al.}, {Properties of plastic scintillators after irradiation },
  Nucl. Instrum. Meth. {\bf{A 552}} (2005) 449--455.

\bibitem{PMT}
{Shi Feng, et al.}, {Performance test of fine-mesh PMT in strong magnetic field
  }, High Energy Phys. Nucl. Phys. (in Chinese) {\bf{28}}~(10) (2004)
  1104--1108.

\bibitem{birks}
{J. B. Birks}, {Scintillations from Organic Crystals: Specific Fluorescence and
  Relative Response to Different Radiations }, Proc. Phys. Soc. {\bf{A 64}}
  (1951) 874.

\bibitem{extrapolation}
{L. L. Wang, et al.}, {BESIII Track Extrapolation and Matching}, High Energy
  Phys. Nucl. Phys. (in Chinese) {\bf{31}} (2007) 183--188.

\bibitem{evtfilter}
{C. D. Fu, et al.}, {Study of the online event filtering algorithm for BESIII},
  Chin. Phys. {\bf{C 32}} (2008) 329--337.

\bibitem{betheformula}
{Particle Data Group (C. Patrignani, et al.)}, {Review of Particle Properties},
  Chin. Phys. {\bf{C 40}}~(10) (2016) 441--455.

\bibitem{landau}
{L. Landau}, {On the energy loss of fast particles by ionization}, J. Phys.
  (USSR) {\bf{8}} (1944) 201--205.

\end{thebibliography}

\end{document}